\documentclass[12pt,preprint]{aastex}

\shorttitle{Accretinng Gas Sphere}
\shortauthors{HANAWA AND SOEDA}

\begin{document}

\title{Critical Accretion Rate for Triggered Star Formation}

\author{Tomoyuki Hanawa\altaffilmark{1}  
and Akihito Soeda\altaffilmark{1}}
\altaffiltext{1}{Center for Frontier Science,
Chiba University, Inage-ku, Chiba, 263-8522, Japan}
\email{hanawa@cfs.chiba-u.ac.jp}

\begin{abstract}
We have reexamined the similarity solution for a self-gravitating
isothermal gas sphere and examined implication to star formation
in a turbulent cloud.  When parameters are adequately chosen, 
the similarity solution expresses an accreting isothermal gas sphere 
bounded by a spherical shock wave.  The mass and radius of the sphere
increases in proportion to the time, while the central density
decreases in proportion to the inverse square of time.
The similarity solution is specified by the accretion rate and 
the infall velocity.  The accretion rate has an upper limit
for a given infall velocity.   When the accretion rate is
below the upper limit, there exist a pair of similarity 
solutions for a given set of the accretion rate and infall
velocity.  One of them is confirmed to be unstable against
a spherical perturbation.  This means that the gas sphere 
collapses to initiate star formation only when the accretion 
rate is larger  than the upper limit.  We have also examined 
stability of the similarity solution against non-spherical
perturbation.  Non-spherical perturbations are found to be
damped.
\end{abstract}


\section{INTRODUCTION}

Similarity solutions have contributed very much to our 
understanding of star formation process.  The classical
similarity solution by \citet{larson69} and \citet{penston69}
elucidated the runaway nature of gravitational collapse.
The density increases in proportion to the inverse square
of the time during the runaway collapse phase.  We learned 
from the similarity solutions of \citet{shu77} and \citet{hunter77} 
that the accretion rate of a protostar is of the order of
$ c _s {}^3 / G $ where $ G $ and $ c _s $ denote the
gravitational constant and the isothermal sound speed of gas.

The similarity solution is also used to evaluate the effects
of rotation and magnetic field.  The similarity solutions of
\citet{narita84} and \citet{saigo98} indicated that the runaway 
collapse cannot be prevented by rotation if once initiated.  
Collapse of a rotating magnetized gas cloud is described
by the similarity solution of \citet{krasnopolsky02}.
Ambipolar diffusion is taken into account in the the 
similarity solution of \citet{adams07}.
\citet{tsai95} extended the similarity solution to include 
shock wave.  \citet{shu02} extended the similarity solution 
involving a shock wave for application to champagne phase
of an \ion{H}{2} region. 

\citet{tsai95} found two classes of similarity solution;
the first class describes accretion onto protostar while the 
second one does failure of star formation.  The central density
decreases in proportion to the inverse square of the
time, $ \rho _c \, \propto \, t ^{-2} $, in the second class
solution.  Although this solution has not gained much 
attention thus far, it provides an insight on dynamical
compression of a molecular cloud core.
If a dense clump of gas is compressed by an external
force, the temporal increase in the density may trigger
gravitational collapse and star formation.  One can surmise
existence of threshold of gravitational collapse.  If the
dynamical compression is either weak or short, the clump
will bounce back to expansion.  A shock wave will be formed
when accreting gas is stoped by the expansion \citep[see, e.g.]
{adams07}.
The similarity solution of 
\cite{tsai95} demonstrated that a spherical cloud can expand
even when it is steadily compressed by a shock wave. 
On the other hand, the shock compressed gas sphere will
collapse owing to its self gravity if the shock is strong
and lasts for a long enough period. 

In this paper we reexamine the similarity solution of 
\cite{tsai95} while keeping its negative implication
in mind.  We find the condition for existence of similarity
solution describing expansion of a gas sphere.  
Conversely it will tell us condition for a shock compressed
gas sphere to collapse by its self gravity.
We also study stability of the similarity solution.
The similarity solution denies collapse due to the self
gravity only when it is stable.

We review the similarity solution in \S 2.1 and show
the method of linear stability analysis in \S 2.2.  
Technical details on the stability analysis are given
in Appendix.  Properties of similarity solutions, such as
accretion rate and infall velocity, are shown in \S 3.1.
Stability of the similarity solution is given in \S 3.2
and \S 3.3. We discuss implications of our analysis in \S 4.

\section{Model and Methods of Computation}

\subsection{Similarity Solution}

We consider an isothermal gas of which distribution is spherically 
symmetric.  Then the hydrodynamical equations are expressed as
\begin{equation} 
\frac{\partial \rho}{\partial t} \; + \;
\frac{1}{r^2} \, \frac{\partial}{\partial r} \, (r^2 \rho v) 
\; = \; 0 \, , \label{hydro1}
\end{equation}
\begin{equation}
\frac{\partial v}{\partial t} \; + \;
v \, \frac{\partial v}{\partial r} \; + \;
\frac{1}{\rho} \, \frac{\partial P}{\partial r} \; + \;
\frac{G M_r}{r ^2} \; = \; 0 \, , \label{hydro2}
\end{equation}
and
\begin{equation}
\frac{\partial M _r}{\partial r} \; = \; 4 \pi r ^2 \, \rho
\label{hydro3}
\end{equation}
where
\begin{equation}
P \; = \; c _{\rm s} ^2 \rho \, .
\end{equation}
Here, the symbols, $ \rho $, $ v$, $ P $, $ M _r $, $ G $, and 
$ c _{\rm s} $ denote the density, velocity, pressure,
mass inside the radius $r $, gravitational constant, 
and the isothermal sound speed, respectively.  
As originally shown by Larson (1969)
and Penston (1969), the hydrodynamical equations have
similarity solutions,
\begin{eqnarray}
\rho (r, \, t) & = & \frac{\varrho (\xi)}{4 \pi G t ^2} 
\, , \label{sim1}\\
v \, (r, \, t) & = & \frac{u (\xi)}{c _s} \, , \label{sim2}\\
M _r (r, \, t) & = & \frac{c _{\rm s} ^3 \, t}{G} \, \mu (\xi) 
\label{sim3}
\end{eqnarray}
where
\begin{equation}
\xi \; = \; \frac{r}{c _{\rm s} \, t} \; .
\label{sim4}
\end{equation}

We restrict ourselves in the case of $ t \, > \, 0 $ in the
following.  Substituting Equations (\ref{sim1}) through (\ref{sim4}) into
Equations (\ref{hydro1}) through (\ref{hydro3}) we obtain
\begin{eqnarray}
\frac{\partial \varrho}{\partial \xi} & = & 
- \, \frac{\left[ \mu \, - \, 2 \, (\xi \, - \, u) ^2 \right] \, \varrho}
{\left[(u \, - \, \xi) ^2 \, - \, 1 \right] \, \xi} \label{sim5}\\
\frac{\partial u}{\partial \xi} & = &
\frac{(\mu \, - \, 2) \, (\xi \, - \, u)}
{\left[(u \, - \, \xi) ^2 \, - \, 1 \right] \, \xi} \label{sim6}
\end{eqnarray}
and
\begin{equation}
\mu \; = \; \xi \, (\xi \, - \, u) \, \varrho \, .
\end{equation}

We assume that the density is finite at the center since
we are interested in application to a starless core, i.e.,
case of no star formation.
Then the density and velocity should be expressed as
\begin{eqnarray}
\varrho & = & \varrho _0 \; - \; \frac{\rho _0}{6}
\left( \varrho _0 \, - \, \frac{2}{3} \right) \, \xi ^2 
\, + \, \frac{\rho _0}{45} \, \left( \varrho _0 \, - \, 
\frac{2}{3} \, \right) \, \left( \varrho _0 \, - \,
\frac{1}{2} \right) \, + \, {\cal O} \, (\xi ^6)  \\
u & = & \frac{2 \xi}{3} \, - \, \frac{\varrho _0}{45} \, 
\left( \varrho _0 \, - \, \frac{2}{3} \right) \, \xi ^3 \, + \, 
{\cal O} \, (\xi ^5) 
\end{eqnarray}
near the center.  Following \citet{tsai95} we assume that
the flow has a shock wave at $ \xi \, = \, \xi _{\rm sh} $.
Then the Rankine-Hugonio relation gives us the condition,
\begin{equation}
\frac{\varrho _+}{\varrho _-} \; = \;
\frac{u _+ \, - \, \xi _{\rm sh}}{u _- \, - \, \xi _{\rm sh}} 
\, , 
\end{equation}
and
\begin{equation}
(u _+ \, - \, \xi _{\rm sh}) \, (u _+ \, - \, \xi _{\rm sh}) 
\; = \; 1 \, .
\end{equation}
where $ \varrho _+ $ and $ \varrho _- $ denote the
the densities of the pre- and post-shocked gases,
respectively, and$ u _+ $ and $ u _- $ do the velocities of them,
respectively. 

In the region far from the origin, a solution of Equations 
(\ref{sim5}) and (\ref{sim6}) approaches the asymptotic solution,
\begin{eqnarray}
\varrho & = & \frac{\dot{M}}{v _{\rm inf} \xi ^2} \; + \;
{\cal O} \, (\xi ^{-3}) \, , \\
u & = & - \, v _{\rm inf} \; + \; {\cal O} \, (\xi ^{-1}) \, .
\end{eqnarray}
The symbol, $ v _{\rm inf}, $ denotes the infall velocity
at the infinity while $ \dot{M} $ denotes the accretion
rate.  
The similarity solution can be specified either by
($ \varrho _c $,~$ \xi _{\rm sh} $) or by 
($ v _{\rm inf}$,~$\dot{M}$).

Note that Larson-Penston solution 
has the same asymptotic form.  Only when
$ v _{\rm inf} $ and $ \dot{M}$ vanish, 
the solution Equations (\ref{sim5})
and (\ref{sim6}) have a different asymptotic form,
\lq \lq plus solutions\rq \rq of \citet{shu77}.

We integrate equations (\ref{sim5}) and (\ref{sim6}) by
the 4th order Runge-Kutta method from $ \xi \, = \, 0 $
for a given set of $ \varrho _c $ and $ \xi _{\rm sh}$
to obtain a similarity solution.  The infall velocity and
accretion rate are obtained numerically as a function of
$ \varrho _c $ and $ \xi _{\rm sh}$.  We also obtain the
mass enclosed in the shock front, 
\begin{equation}
M _c \; = \; \int _0 ^{\xi _{\rm sh}}  \varrho \,
\xi ^2 \, d \xi \, . 
\end{equation}

\subsection{Spherical and Non-spherical Perturbations}

We have performed a normal mode analysis to examine the stability
of the similarity solution.  In the analysis,
the density is assumed to be expressed as
\begin{equation}
\rho \, (r,~t) \; = \; 
\frac{\varrho (\xi) \, + \, t ^\sigma \,
\delta \varrho (\xi) \, Y _{\ell} ^m \, (\theta,~\varphi)}
{4 \pi G t ^2} \, , \label{density-p}
\end{equation}
where $ Y _{\ell} ^m \, (\theta,~\varphi)$ denotes the
spherical harmonic function.  This particular form is
chosen because the similarity solution has no specific
timescale. An eigenmode has a growth timescale and 
an unstable perturbation grows exponentially, 
when the unperturbed state is stationary and has a 
specific timescale.  When the physical quantities 
vary according to a power law in time in an unperturbed state,
the eigenmode grows (or decay) in proportion to a power of
time, $ | t | ^\sigma$.  See \citet{hanawa99} for the
justification of Equation (\ref{density-p}).  They analyzed
linear stability of the Larson-Penston solution against
a non-spherical perturbation, using the coordinates,
$ (\xi,~\theta,~\varphi,~\ln~t)$ instead of 
the ordinary spherical coordinates, $(r,~\theta,~\varphi,~t)$.
It is shown that the similarity solution can be expressed
as a steady state and an unstable mode grows in propotion
to $ \exp~(\sigma~\ln~t) \, = \, t ^\sigma $ in the coodinates.

The power index, $ \sigma $,
is obtained as an eigenvalue of the perturbation equations
as shown in Appendix.  When the real part of 
$ \sigma $ is positive, the mode grows in time and the
similarity solution is unstable.  We call the real part
of $\sigma$, the growth index in the following according
to the suggestion by F.~H. Shu, the referee of this paper. 
When $ \sigma $ is complex, the mode grows (or decays) while
oscillating.  The imaginary part of $ \sigma $ is related with
the frequency of oscillation, although the physical 
oscillation period increases with time.

\section{Results}

\subsection{Similarity Solutions}

\begin{figure}[!hp]
\centering
\plotone{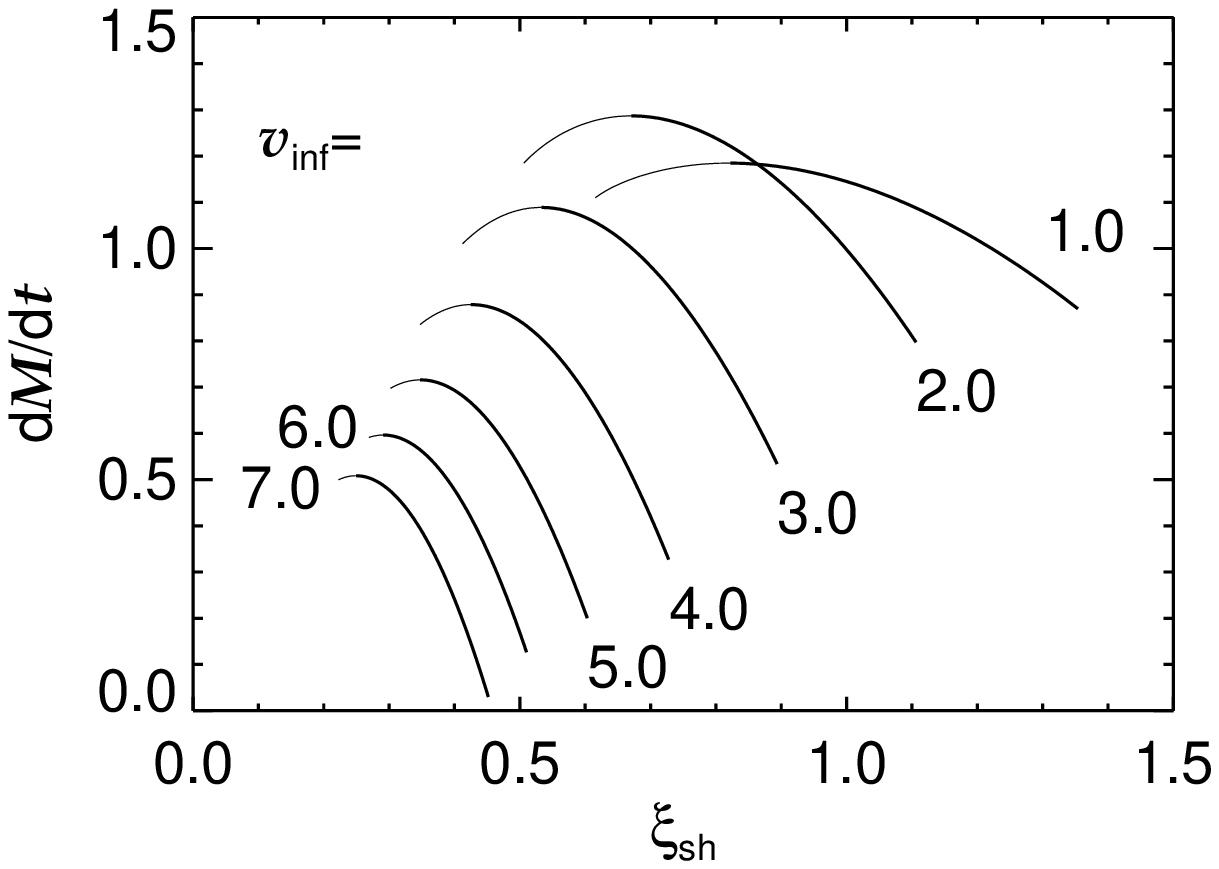}
\caption{The accretion rate, $ \dot{M}$, is shown as a function
of $ xi _{\rm sh}$  for a given $ v _{\rm inf}$.
\label{dMdt}}
\end{figure}

First we obtained a series of similarity solutions for a given
$ \varrho _c $ by increasing $ \xi _{\rm sh} $.  
The infall velocity, $ v _{\rm inf} $, decreases 
monotonically with increase in $ \xi _{\rm sh}$.
It reaches $ v _{\rm inf} \, = \, 0 $ at some $ \xi _{\rm sh} $
and the similarity solution terminates.  By compiling 
the similarity solutions, we obtained $ \dot{M} $ as a function
of $ v _{\rm inf} $ and $ \xi _{\rm sh} $.  
The curves denote $ \dot{M} $ as a function of
$ \xi _{\rm sh} $ for given $ v _{\rm inf} $ in Figure \ref{dMdt}.  

The accretion rate, $ \dot{M} $, is maximum at a certain 
$ \xi _{\rm sh}$ for a given $ v _{\rm inf}$.  It increases
with increase in $ \xi _{\rm sh}$ at a small $ \xi _{\rm sh} $
while it decreases with increase in $ \xi _{\rm sh} $ at a
large $ \xi _{\rm sh}$.  The former is denoted by 
thin curves while the latter is by thick ones.  This means
that there exists two similarity solutions for a given set
of $ v _{\rm inf}$  and $ \dot{M}$. 

\begin{figure}[!hp]
\centering
\plotone{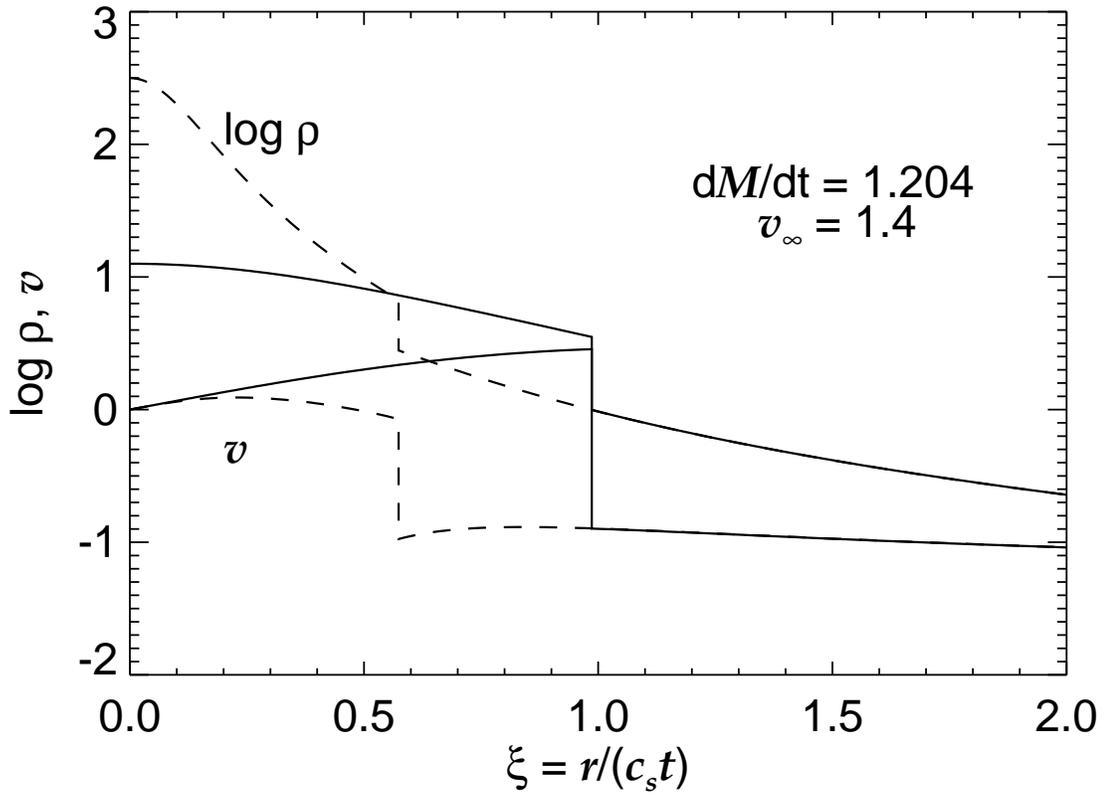}
\caption{The density and velocity distributions are
shown for two similarity solutions having the same $ v _{\rm inf} $
and $ dM/dt $.\label{Solution}}
\end{figure}

Figure \ref{Solution} shows two similarity solutions having
$ v _{\rm inf} \, = \, 1.4 $ and $ \dot{M} \, = \, 1.204 $.
The solid curves denote the solution of 
$ \xi _{\rm sh} \, = \, 0.986 $ while the dashed curves do
that of $ \xi _{\rm sh} \, = \, 0.574 $.  Both the solutions have
the same density and velocity distributions in the region
of $ \xi \, \ge \, 0.986 $.  The main difference is the 
location of the shock front.  The shock compressed gas
sphere is denser and expands more slowly in the solution
denoted by the dashed curve. Since the expansion is slow,
the solution is similar to the \citet{bonnor56}-\citet{ebert55}
solution for a self-gravitating isothermal gas sphere.  
As well as the Bonnor-Ebert sphere, the latter solution
is shown to be unstable (\S 3). 

\begin{figure}[!hp]
\centering
\plotone{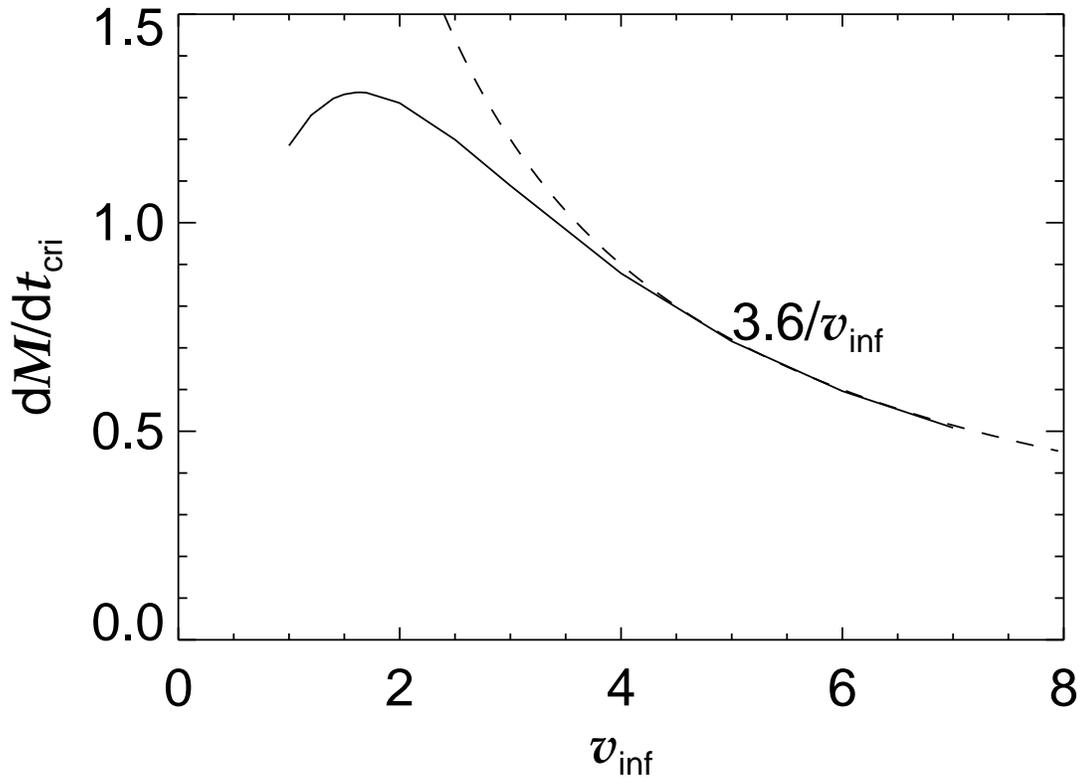}
\caption{The critical accretion rate is shown as a function of
$ v _{\rm inf} $.\label{dMdt-max}}
\end{figure}

Figure \ref{dMdt-max} indicates that the similarity solution has
an upper limit on $ \dot{M} $ for a given $ v _{\rm inf} $.
The upper limit is highest $ \dot{M} _{\rm max} \, = \, 1.312 $ 
at $ v _{\rm inf} \, = \, 1.64 $ .
Similarity solutions do not exist for $ \dot{M} \, > \, 1.312$.
The implication of non-existence is discussed in \S 4.

\subsection{Spherical Perturbations}

\begin{figure}[!hp]
\centering
\plotone{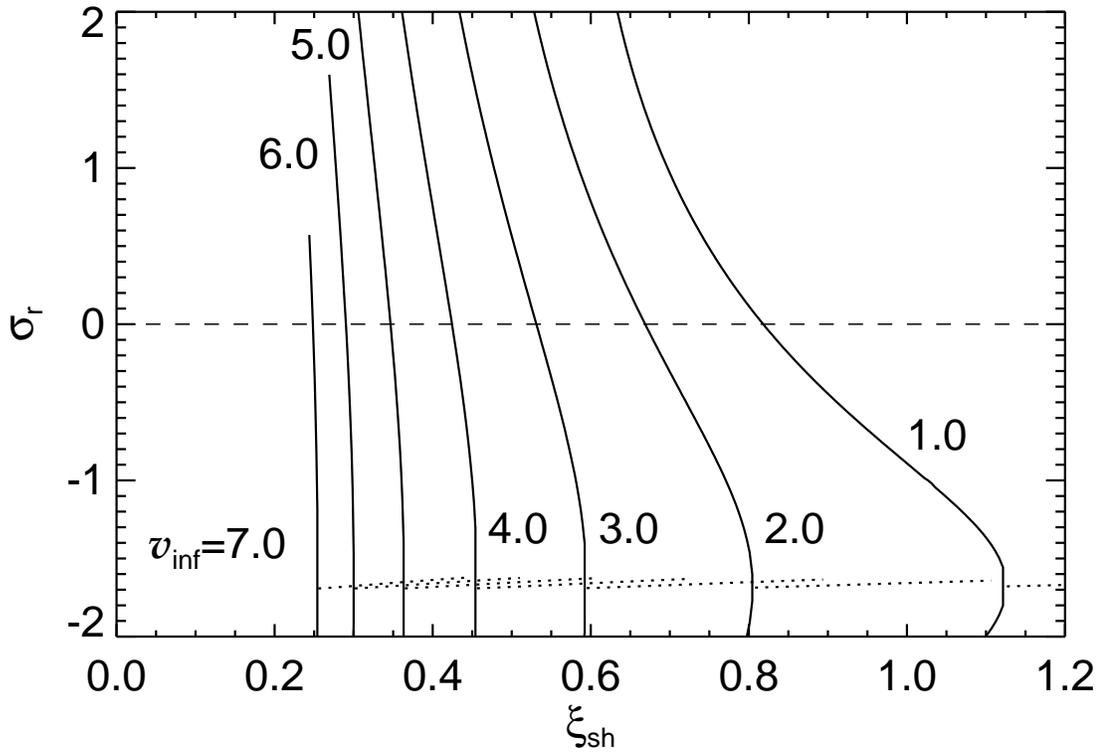}
\caption{Each curve denotes the real part of eigenfrequency, 
$ \sigma _r $, of a spherical perturbation as a function of the
shock radius, $ \xi _{\rm sh} $ for a given $ v _{\rm inf} $.
\label{l=0}}
\end{figure}

Figure \ref{l=0} shows the growth index of the spherical
perturbation, $ \sigma _r $, as a function of $ \xi _{\rm sh} $ for a
series of similarity solutions having a given $ v _{\rm inf}$.
The solid curves denote the growth index of modes having
real index.  The dashed lines denote the real part of
complex indecies.  One of the index is positive
($ \sigma _r \, > \, 0$) and the similarity solution is unstable
only when the shock radius ($ \xi _{\rm sh} $) is smaller than a 
critical value.  The condition of neutral stability coincides with
that of maximum accretion rate for a given $ v _{\rm inf}$ as
expected.  When $ \xi _{\rm sh} $ is a little larger than
the critical, the solution is stable and the most slowly damping
mode has a real index.  When $ \xi _{\rm sh} $ is large enough,
the solution is stable and the most slowly damping mode has 
a complex index.

Our survey is limited to modes having low indecies, i.e.,
in the range $ |\sigma _i | \, \le \, 1.0 $. It is however 
unlikely that we have missed unstable spherical perturbations.
A spherical perturbation induces only sound waves and they are
confined within the gas sphere since it is bounded by the 
shock wave.  A high frequency sound wave has a shorter wavelength
and is unlikely to be Jeans unstable.

\subsection{$ \ell $~=~2 Mode}

We have studied $ \ell $~=~2 mode as a typical non-spherical
perturbation.  This is in part because the dipole ($\ell \, = \, 1 $)
mode is unlikely to be excited.  If the dipole mode grows, the
inner and outer parts of the gas sphere should move in the
opposite direction each other to keep the center of gravity.

\begin{figure}[!hp]
\centering
\plotone{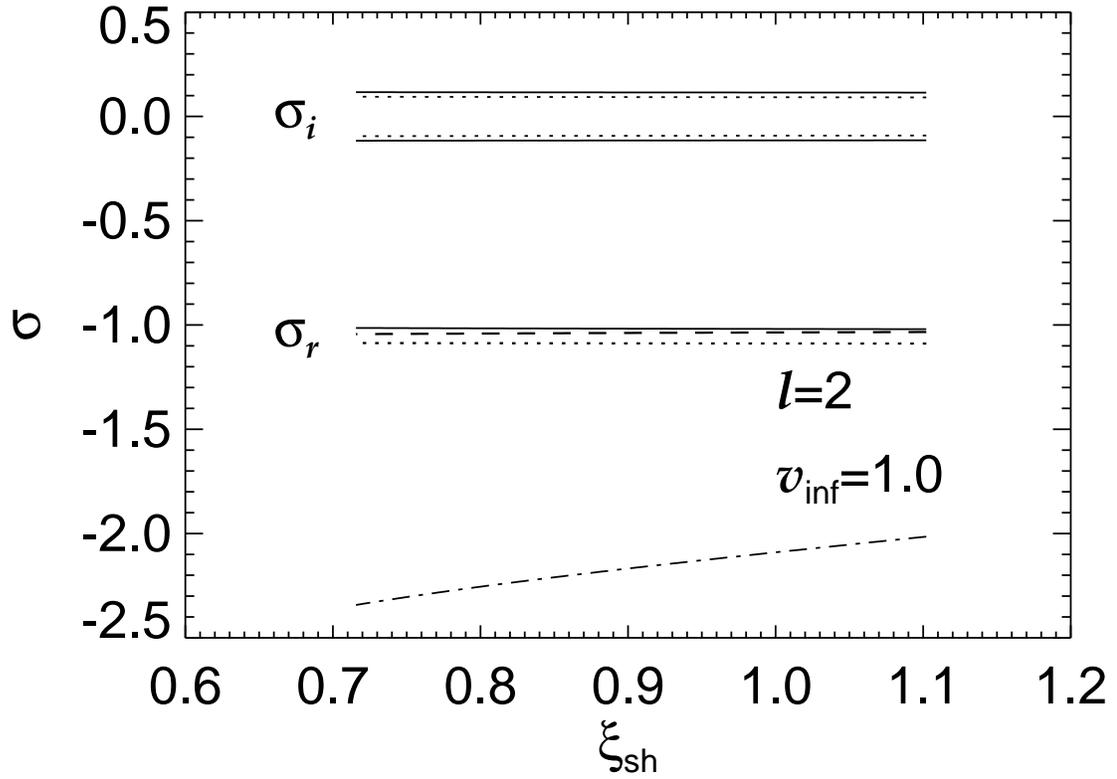}
\caption{The eigenfrequency, $ \sigma $, is shown as a
function of $ \xi _{\rm sh} $ for similarity solutions having
$ v _{\rm inf} \, = \, 1.0 $.  The solid curves denote the
real part of $\sigma $ for non-spherical perturbations
having $ \ell \, = \, 2 $, while dashed curves do that of
imaginary part.  Mode a has complex eigenfrequencies while
modes b and c have real eigenfrequencies.\label{l=2v=1}}
\end{figure}

Figure~\ref{l=2v=1} denotes the eigenfrequencies of the
$ \ell \, = \, 2 $ modes for 
similarity solutions of $ v _{\rm inf} $~=~1.0.
The abscissa denotes $ \xi _{\rm sh } $ while the ordinate
denotes the index, $ \sigma $.  All the modes are
damping ($ \sigma _i \, < \, -1 $).  The solid lines denote
the mode having the smallest damping index.  The mode has a
small imaginary part ($\sigma _i \, \simeq \, \pm 0.1 $).
The mode having the second smallest damping index is denoted by 
the dashed line in Figure~\ref{l=2v=1} and has a real
eigenfrequency (pure damping).  The mode having the third 
smallest damping index has an imaginary part in the 
eigenfrequency.  The imaginary part is similar to that
of the mode having the smallest damping index.  These
three modes have a similar damping index of 
$ -1.1 \, \le \, \sigma _r \, \le \, -1.0 $.  The other
mode (dash-dotted line in Fig.~\ref{l=2v=1}) has a much
larger damping index. 

\begin{figure}[!hp]
\centering
\plotone{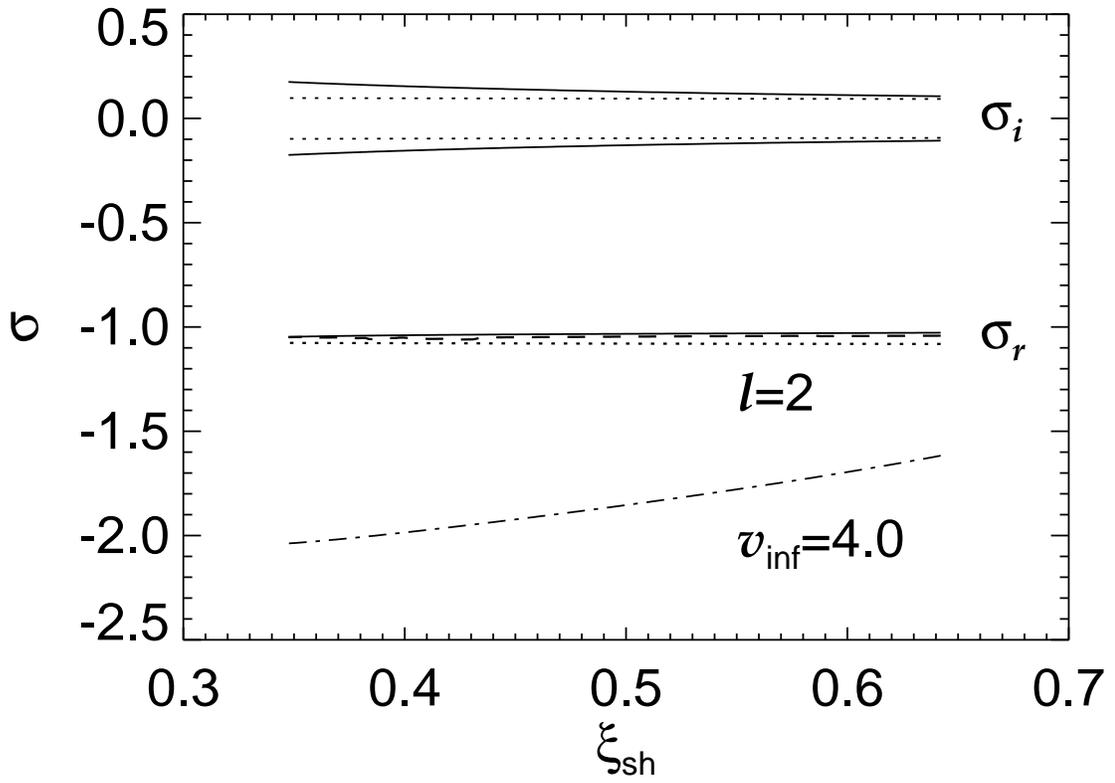}
\caption{The same as Fig,~\ref{l=2v=1} but for 
$ v _{\rm inf} $~=~4.\label{l=2v=4}}
\end{figure}

Figure~\ref{l=2v=4} is the same as Figure~\ref{l=2v=1} but
for $ v _{\rm inf} $~=~4.0.  Again, all the modes are damping.
The damping index is larger than unity ($ \sigma _r \, < \, -1 $).
The oscillation frequency of the smallest damping mode is 
larger than those of $ v _{\rm inf} $~=~1.0.  

One might ask the reason why the similarity solution is
stable against a bar ($\ell~=~2$) mode.  We think that 
the bar mode is damped by expansion of the gas sphere.
Since the radius of the shock front increases with the time, 
the asphericity of the shock compressed gas sphere 
decreases unless the displacement grows faster than 
the radius.  When a gas sphere is collapse, the bar
mode can be excited as shown by \citet{lin65} for the
pressure less gas and \citet{hanawa99} for an isothermal gas.

We have not yet studied the modes of $ \ell \, \ge \, 3 $.  
However they are also unlikely to be unstable, since 
the self-gravity does not excite a short wave perturbation.

\section{Discussion}

We obtained the critical accretion rate above which there 
exits no similarity solution.  The critical rate can be
interpreted as the minimum accretion rate for a high
density clump to initiate self-gravitational collapse.
The critical accretion rate can be rewritten as
\begin{equation}
\left. \frac{dM}{dt} \right| _{\rm cr} \; = \;
\frac{3.6 \, c _s ^4}{G v} \, , \label{critical-phys}
\end{equation}
for $ v \, \ga \, 3 c _s$ in the dimensional form.

Equation (\ref{critical-phys}) gives us an estimate
for a converging flow to initiate gravitational
collapse.  We shall consider a spherical region of
which surface is surrounded by a converging flow. 
The radius, inflow velocity, and density are assumed
to be $ r$, $ v $ and $ \rho $, respectively.  Then
the gravitational collapse will be initiated when
the mass accretion rate exceeds the critical,
\begin{equation}
4 \pi r ^2 \, \rho \, v \; > \; 
\frac{3.6 \, c _s ^4}{G \, v} \, . \label{Jeans1}
\end{equation} 
Equation (\ref{Jeans1}) can be rewritten as
\begin{equation}
\frac{2r}{\lambda _{\rm J}} \; > \; 
\frac{0.3 \, c _{\rm s}}{v} \, , \label{Jeans2}
\end{equation}
where
\begin{equation}
\lambda _{\rm J} \; = \; \frac{2 \pi \, c _s}
{\displaystyle \sqrt{4 \pi G \rho}} \; .
\end{equation}
Since $ \lambda _{\rm J} $ denotes the Jeans length,
Equation (\ref{Jeans2}) means that the effective Jeans
length reduces in proportion to the inverse of the
Mach number. 

The Jeans mass is proportional to the cube of the Jeans length 
for a given density.  Thus the effective Jeans mass
should reduce to
\begin{eqnarray}
M _{\rm J,~eff} & = & M _{\rm J} \, \left( 
\frac{\lambda _{\rm J,~eff}}{\lambda _{\rm J}}\right) ^3 \\
& = & M _{\rm J} \, \left( \frac{v}{0.3~c_s}\right) ^{-3} \, .
\end{eqnarray}
This implies that compression of sub Jeans mass clump may
result in gravitational collapse in the region of flow 
convergence.  Note that the effective Jeans mass is
several order of magnitude smaller than the classical
one when $ v \, \ga \, 3 \, c _s $.

The compression should continue for a certain timescale for
a dynamically compressed clump to collapse by its self gravity.
If we evaluate the minimum timescale to be the effective Jeans length
divided by the flow velocity, it is shorter than the free-fall
timescale by a factor of the Mach number squared,
\begin{equation}
\tau _{\rm comp} \; \simeq \; \frac{\lambda _{\rm J,~eff}}{v} \;
\simeq \; \tau _{\rm ff} \, \left( \frac{v}{c _s}\right) ^{-2} \, .
\end{equation}

The timescale can be translated into the wavelength of perturbation.
A compressed clump can collapse by the self gravity if the wavelength 
of velocity perturbation is longer than the effective Jeans length.  
If turbulence contains velocity perturbations of long wavelengths,
gravitational collapse due to dynamical compression will take place
somewhere in the cloud. In such case we can expect a number of
clumps of which masses are much smaller than the classical Jeans mass.

\acknowledgments

We thank T. Matsumoto and F. Nakamura for discussion and 
valuable comments on the original manuscript.  
We also thank F.~H. Shu for his valuable comments as
the referee. 
We have added comparison of 
\citet{tsai95} solution with Bonnor-Ebert
solution in the revised manuscript according to the comments
given to the original manuscript.  
This study is financially supported in part by the 
Grant-in-Aid for Scientific Research on Priority Area (19015003) of 
The Ministry of Education, Culture, Sports, Science, and
Technology (MEXT).

\appendix

\section{Linear Stability Analysis}

Since the density is given by Equation (\ref{density-p})
in our linear stability analysis, the gravitational potential 
should be expressed as
\begin{equation}
\Phi \,(r,~t) \; = \; c _s ^2 \, \left[
\psi \, (\xi) \; + \; t ^\sigma \, \delta \psi (\xi) \,
Y _{\ell} ^m (\theta,~\varphi) \right] \, . 
\end{equation}
The velocity, $ \mbox{\boldmath$v$} \, = \, 
(v _r,~v_\theta,~v_\varphi) $,
is assumed to be expressed as
\begin{eqnarray}
v _r (r,~t) & = & c _s \, \left[ u (\xi) \, + \, t ^\sigma \, 
\delta u _r \, Y _{\ell} ^m (\theta,~\varphi) \right] \, ,\\
v _\theta (r,~t) & = &  - \, c _s \, t ^\sigma \,
\frac{\delta u _\theta (\xi)}{\ell \, + \, 1} \,
\frac{\partial}{\partial \theta}
Y _\ell ^m (\theta,~\varphi) \, , \\ 
v _\varphi (r,~t) & = &  - \, c _s \, t ^\sigma \,
\frac{\delta u _\theta (\xi)}{(\ell \, + \, 1) \, \sin \theta} \,
\frac{\partial}{\partial \varphi}
Y _\ell ^m (\theta,~\varphi) \, , 
\end{eqnarray}
in the spherical coordiantes.
We assume that the perturbation is vanishingly small in the
region very far from the center.  In other words, we restrict
ourselves to search for an instability due to the
internal structure of the shock compressed gas sphere.
Thus the flow is assumed to have no
vorticity,
\begin{equation}
\left( \ell \, + \, 1 \right) \, \delta u _r \; + \;
\frac{d}{d\xi} \, \left(\xi \, \delta u _\theta \right)
\; = \; 0 \, .
\end{equation}

The assumption of no vorticity is based on the Kelvin's
circulation theorem.
It says the circulation,
\begin{equation}
\Gamma \; \equiv \; \oint _{C} \mbox{\boldmath$u$}\cdot
d\mbox{\boldmath$u$}
\end{equation}
is conserved for an isothermal (barotropic) gas
\citep[see, e.g., Chapter 6 of][]{shu92}.
Since the Lagrangian loop {\it C} expands, the
circulation velocity should decrease with the time 
to conserve $ \Gamma $, although it can grow
with the time if measured at a given $ \xi $.
See \citet{hanawa00} for more details on the mathematical
treatment of perturbations having vorticity.

After some manipulation, the perturbation equations are 
written as 
\begin{equation}
\left(\sigma \, + \, 1 \right) \, \delta \varrho \; + \;
\frac{1}{\xi ^2} \, \frac{~d}{d\xi} \,
\left( \xi ^2 \, \delta f \right) \; + \; 
\frac{\ell \, \varrho _0 \, \delta u _\theta}{\xi} \; = \; 0 \, ,
\label{delta-rho}
\end{equation}
\begin{eqnarray}
\left(\sigma \, + \, 2 \right) \, \delta f & + &
\frac{1}{\xi ^2} \, \left\{ \xi ^2 \,
\left[ 2 w \, \delta f \, + \, (1 \, - \, w ^2) \right]
\right\} \nonumber \\
& \; & - \, \left(\varrho _0 w \, + \, \frac{2}{\xi}\right)
\, \delta \varrho \; + \; \varrho _0 \delta \gamma
\; + \; \frac{\ell \varrho _0 w \delta u _\theta}{\xi} \;
= \; 0 \, ,
\end{eqnarray}
\begin{equation}
\frac{~d}{d\xi} \, \delta \psi \; = \;
\delta \gamma \, ,
\end{equation}
\begin{equation}
\frac{~d}{d\xi} \, \delta \gamma \; = \;
\delta \varrho \; - \; \frac{2 \, \delta \gamma}{\xi} 
\; - \; \frac{\ell \, (\ell \, + \, 1)}{\xi ^2} \,
\delta \psi \label{delta-gamma}
\end{equation}
where
\begin{eqnarray}
w & = & u \, - \, \xi \, , \\
\delta u _\theta & = &
\frac{\ell \, + \, 1}{(\sigma \, + \, 1) \, \xi} \, 
\left( w \, \delta u _r \; + \; \frac{\delta \varrho}{\varrho _0} 
\; + \; \delta \psi \right) \, , \\
\delta f & = & \varrho _0 \, \delta u _r \; + \; w \,
\delta \varrho \, ,\\
\delta \gamma & = & \frac{\partial}{\partial \xi} \, 
\delta \psi \, .
\end{eqnarray}
The perturbation equations do not contain the azimuthal
wavenumber, $ m $, and accordingly thus the growth rate 
does not depend on $ m $.  This is because the unperturbed
state (similarity solution) is spherically symmetric.

We must consider the jump condition at the shock front
as well as the boundary conditions to solve the perturbation
equation.  The density perturbation is discontinuous
between the pre- and post-shocked flows and contains 
change due to shift of the shock front.  The latter 
introduces the Dirac's delta function proportional
to the shift and the density difference between the 
pre- and post flows.  After some manipulation we obtain
the jump condition,
\begin{eqnarray}
\left(w _+ \, - \, w _-\right) \, \left[ \delta f \right]
\; - \; \left(\varrho _+ \, - \, \varrho _- \right) \,
\left[w \delta u _r \, + \, \frac{\delta \varrho}{\varrho _0} \right] 
& = & 0 \, , \\
\left(\sigma \, + \, 1 \right) \, \left[\delta \gamma \right] \;
+ \; \left[ \delta f \right] & = & 0 \, 
\end{eqnarray}
where the bracket denotes the difference between 
$ \xi \, = \, \xi _{\rm sh} \, \pm \, \varepsilon $.
The symbols with the subscripts, $ + $ and $ - $, denote the 
values at $ \xi \, = \, \xi _{\rm sh} \, \pm \, \varepsilon $,
respectively. The perturbation in the potential, $ \delta \psi $,
is continuous even at $ \xi \, = \, \xi _{\rm sh}$.
These jump conditions enable us to connect the perturbation
in the post-shocked flow ($ \xi \, < \, \xi _{\rm sh}$) and
that in the pre-shocked flow ($\xi \, > \, \xi _{\rm sh}$).

When the perturbation is spherically symmetric ($ \ell \, = \, 0 $),
Equations (\ref{delta-rho}) and (\ref{delta-gamma}) are linearly
dependent and we obtain
\begin{equation}
\delta \gamma \; = \; - \, \frac{\delta f}{\sigma \, + \, 1}
\end{equation}
for $ \ell \, = \, 0$.  Thus we need two independent boundary
conditions for $ \ell \, = \, 0 $ and three for $ \ell \, \ne \, 0 $.

The spherical perturbation should vanish in the pre-shocked
since the unperturbed flow is supersonic.  Otherwise the
perturbation diverges when the mode is unstable. 
Thus the effective outer boundary is set at $ \xi \, = \,
\xi _{\rm sh} $ and the jump condition is applied there.
The other boundary condition is set so that the velocity
perturbation vanishes at the origin, $ \xi \, = \, 0 $
and is proportional to $ \xi $ near the origin.
We have integrated the perturbation equation numerically
with the Runge-Kutta method from the
origin and searched the eigenvalue, $ \sigma $, by
try and error. 

Non-spherical perturbations should also be regular at the origin
and vanishingly small at infinity.  We calculated asymptotic solutions
around the origin and those around the infinity to examine the 
condition for perturbations to be regular according to 
\citet{hanawa99}.   
Some asymptotic solutions are divergent since the perturbation
equations are singular at the origin and infinity.
When $ \ell \, \ne \, 0 $, the perturbation should be
expressed by the asymptotic solution,
\begin{eqnarray}
\delta \varrho & = & B \varrho _0 \, \xi ^\ell \, , 
\label{delta-rho-0}\\
\delta u _r & = & - \, A \ell \xi ^{\ell \, - \, 1}
\; - \; C \, (\ell \, + \, 2) \, \xi ^{\ell \, + \, 1} \, , 
\label{delta-u-0}\\
\delta \psi & =& \left[A \, \left(\sigma \, + \, 1 \, - \,
\frac{\ell}{3} \right) \, - \, B\right] \, \xi ^\ell \, ,
\label{delta-psi-0}
\end{eqnarray}
where 
\begin{equation}
C \; = \; \frac{1}{4 \ell \, + \, 6} \,
\left[ \frac{\ell}{3} \, \left(\varrho _0 \, - \, \frac{2}{3} 
\right)\, A \; + \; \left(\sigma \, - \, \frac{\ell \,- \, 1}{3} 
\right)\, B \right] \, .
\end{equation}
Here the symbols, $ A $ and $ B $, denote arbitrary constants.
Equations (\ref{delta-rho-0}) and (\ref{delta-psi-0}) mean that 
the density and potential perturbations can be expanded by 
a series of polynomials starting from the $\ell$-th power 
in the Cartesian coordinates.  Otherwise the perturbation 
diverges at the origin.  The velocity perturbation is also
expanded by another series of polynomials starting from 
the $ (\ell - 1) $-th order since it should be balanced
with gradient of the potential perturbation.

The other boundary condition is given by the asymptotic
solution,
\begin{eqnarray}
\delta \varrho & = & \frac{2 D \, (\ell \, + \, 1) \, \varrho _0}
{(\sigma \, + \, \ell \, + \, 2)(\sigma \, + \, \ell \, + \, 3) \,
\xi ^{\ell \, + \, 3}} \, , 
\label{delta-rho-inf}
\\
\delta u _r & = & \frac{D \, (\ell \, + \, 1)}
{(\sigma \, + \, \ell \, + \, 2) \, \xi ^{\ell \, + \,2}} \, , 
\label{delta-u-inf}
\\
\delta \psi & = & \frac{D}{\xi ^{\ell\, + \,1}} \, ,
\label{delta-psi-inf}
\end{eqnarray}
for $ \xi \, \gg \, 1$.  Here, the symbol $D $ denotes an
arbitrary constant. Equation (\ref{delta-psi-inf}) means
that the potential perturbation in the region 
$ \xi \, \gg \, 1 $ is dominated by the 
aspherical density distribution near the center.
In other words the density perturbation is too small to
affect the gravitational potential. Equation (\ref{delta-u-inf})
means that the velocity perturbation is induced by the gravitational
perturbation.  Equation (\ref{delta-rho-inf}) means that
the density perturbation is induced by the velocity perturbation.
Thus our boundary conditions gurantee that the perturbation
is induced by internal change in the flow.

We integrated the perturbation equations
both from the origin ($ \xi \, = \, 10 ^{-2} $)
and a very large $ \xi \,(\simeq \, 100)$
with the Runge-Kutta method. 
We surveyed the eigenvalue, $ \sigma $, by examining whether
the numerically solutions from both ends satisfy the jump
condition at the shock front.


\clearpage

\begin{figure}[!hp]
\centering
\plotone{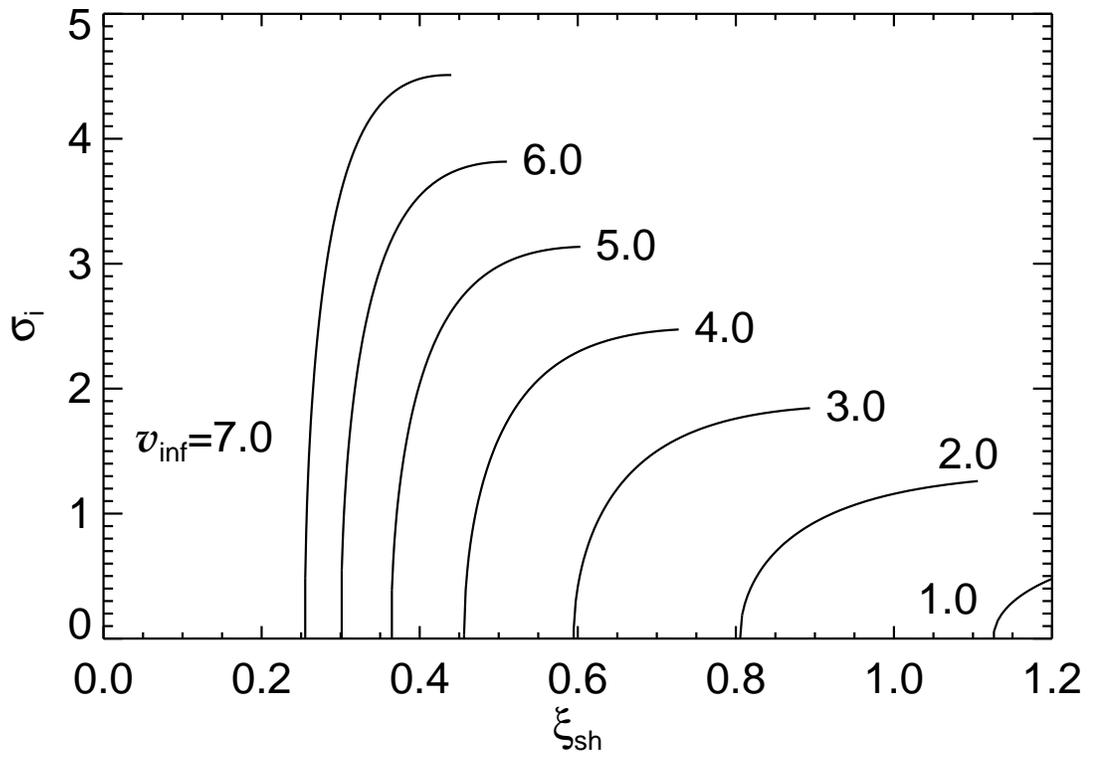}
\caption{The same as Fig.~\ref{l=0} but for the imaginary
part, $ \sigma _i $.\label{l=0imag}}
\end{figure}

\end{document}